\documentclass[prb,amsmath,twocolumn,amssymb,showpacs]{revtex4}
\usepackage{graphicx}
\usepackage{bm}
\usepackage{color}
\begin{document}
\title{Dirac cones beyond the honeycomb lattice: a symmetry based approach}
\author{Guido van Miert and Cristiane Morais Smith}

\affiliation{Institute for Theoretical Physics, Centre for Extreme Matter and Emergent Phenomena, Utrecht University, Leuvenlaan 4, 3584 CE Utrecht, The Netherlands}

\begin{abstract}
Recently, several new materials exhibiting massless Dirac fermions have been proposed. However, many of these do not have the typical graphene honeycomb lattice, which is often associated with Dirac cones. Here, we present a classification of these different two-dimensional Dirac systems based on the space groups, and discuss our findings within the context of a minimal two-band model. In particular, we show that the emergence of massless Dirac fermions can be attributed to the mirror symmetries of the materials. Moreover, we uncover several novel Dirac systems that have up to twelve inequivalent Dirac cones, and show that these can be realized in (twisted) bilayers. Hereby, we obtain systems with an emergent SU(2N) valley symmetry with N=1,2,4,6,8,12. Our results pave the way to engineer different Dirac systems, besides providing a simple and unified description of materials ranging from square- and $\beta$-graphynes, to Pmmn-Boron, TiB$_2$, phosphorene, and anisotropic graphene.
\end{abstract} 
\maketitle
\section{Introduction}
The synthesis of graphene has opened the world of Dirac cones and the physics that derives from them [\onlinecite{RMP-CastroNeto2009}]. Since then, Dirac systems have attracted a huge interest not only from a fundamental point of view, but also with a view on applications. Indeed, graphene holds promises to revolutionize nanotechnology because it is ultrathin, light, transparent, and resilient to bending. Current applications range from sensors to transparent flexible electronics, field-effect devices, and spintronics. In addition, many phenomena first discussed in the realm of elementary-particle physics now find their way to tabletop experiments. Examples include the Klein paradox, Zitterbewegung, and connections to gravity [\onlinecite{NatPhys2006}]. Moreover, Dirac fermions have been of great importance to the field of topological quantum matter. Firstly, experiments at very high magnetic fields revealed that in graphene the quantum Hall effect may be observed at room-temperature.\cite{IQHE} Secondly, graphene is the system for which the quantum spin Hall effect was originally proposed.\cite{Kane-Mele-PRL2005}

During the first decade after the initial synthesis of graphene, the field of Dirac materials has evolved into various directions. One of these focuses on graphene's honeycomb lattice, and is concerned whether one can obtain different or more versatile materials by replacing the carbon atoms by different group IV elements [\onlinecite{RSC-Heine}]. This pursuit led to the successful synthesis of silicene, germanene, and stanene.\cite{silicene,germanene,stanene} Since these atoms have a larger atomic radius than carbon, they buckle out of the two-dimensional plane and this buckling is useful for applications in electronic devices. In particular, one can generate a mass for the Dirac fermions in silicene by the application of an external electric field, which is exploited in field-effect transistors.\cite{FET} Moreover, stanene is supposedly a near room-temperature topological insulator.\cite{RTTI} Furthermore, honeycomb lattices composed of alternating Boron (Silicon) and Nitrogen (Carbon) atoms have been synthesized.\cite{BN,SiC} In these binary compounds, the asymmetry between the different atoms induces a large gap in the spectrum, such that SiC is a semiconductor and hBN is an insulator.

Although the honeycomb lattice has proven to be a very successful platform for designing novel Dirac materials, quite some progress has been made in the field of non-honeycomb systems as well. A very promising class of materials in this respect is provided by graphynes.\cite{malko2012} These are two-dimensional carbon allotropes that are characterized by the presence of acetylene bonds ($-C\equiv C-$) in their lattice structure. In particular, the alternating presence and absence of the acetylene bond leads to more complicated lattice structures, and consequently to a richer band structure. For example, $\beta-$graphyne exhibits six inequivalent Dirac cones in the Brillouin zone (BZ), whereas the recently proposed square grahynes have four.\cite{sgraphyne}   Although graphynes have not yet been experimentally realized, the structurally similar graphdiyne has already been successfully synthesized.\cite{Li2010} Recent studies show that similar Dirac systems can also be realized in non-carbon materials. Indeed, a monolayer of the metal diboride TiB$_2$ exhibits six Dirac cones,\cite{TiB2} and the rectangular Boron allotrope pmmn Boron has one pair of Dirac cones.\cite{pmmnB}

The above examples show that Dirac materials can thus have very different crystalline lattices, and may even be composed of different elements. If we forget in a first instance about the chemical composition, and focus instead on symmetries, we may distinguish three different classes of Dirac materials. The first class is formed by graphene, and the other materials that have the Dirac cones located at the high-symmetry (HS) K and K' points. Similarly, we may group together the Dirac systems that have their cones located along the HS lines. $\beta-$graphyne, square graphynes, pmmn Boron, and TiB$_2$ belong to this class. Finally, one may consider the systems that have their Dirac cones located at generic points in the BZ. A typical example of those is the organic conductor $\alpha$ -(BEDT-TTF)$_2$I$_3$.\cite{ETI}
 
Given the diversity of chemical composition and lattice structure of the current Dirac materials, it is essential to acquire a better understanding of the conditions for the emergence of Dirac cones. This is precisely the aim of this paper. Therefore, we first discuss in Sec.~II the occurrence of band-crossing points in two-dimensional crystals. Then, in Sec.~III, we show that the Dirac cones along HS lines are a mere consequence of the mirror-reflection symmetry of the different materials. This allows us to promptly engineer Hamiltonians that display Dirac behaviour, and in particular, to realize a multitude of Dirac cones, which bring us from the SU(4) case in graphene to a generic SU(2N), for N-valley degrees of freedom (N=1,2,4,6,8,12). Then, in Sec.~IV we extend the discussion to systems that have Dirac cones located at arbitrary points in the BZ. Throughout this paper, we illustrate our analysis by discussing the various cousins of graphene. Our conclusions are presented in Sec.~V.

\begin{figure*}[t]
\centering

\includegraphics[width=\textwidth]{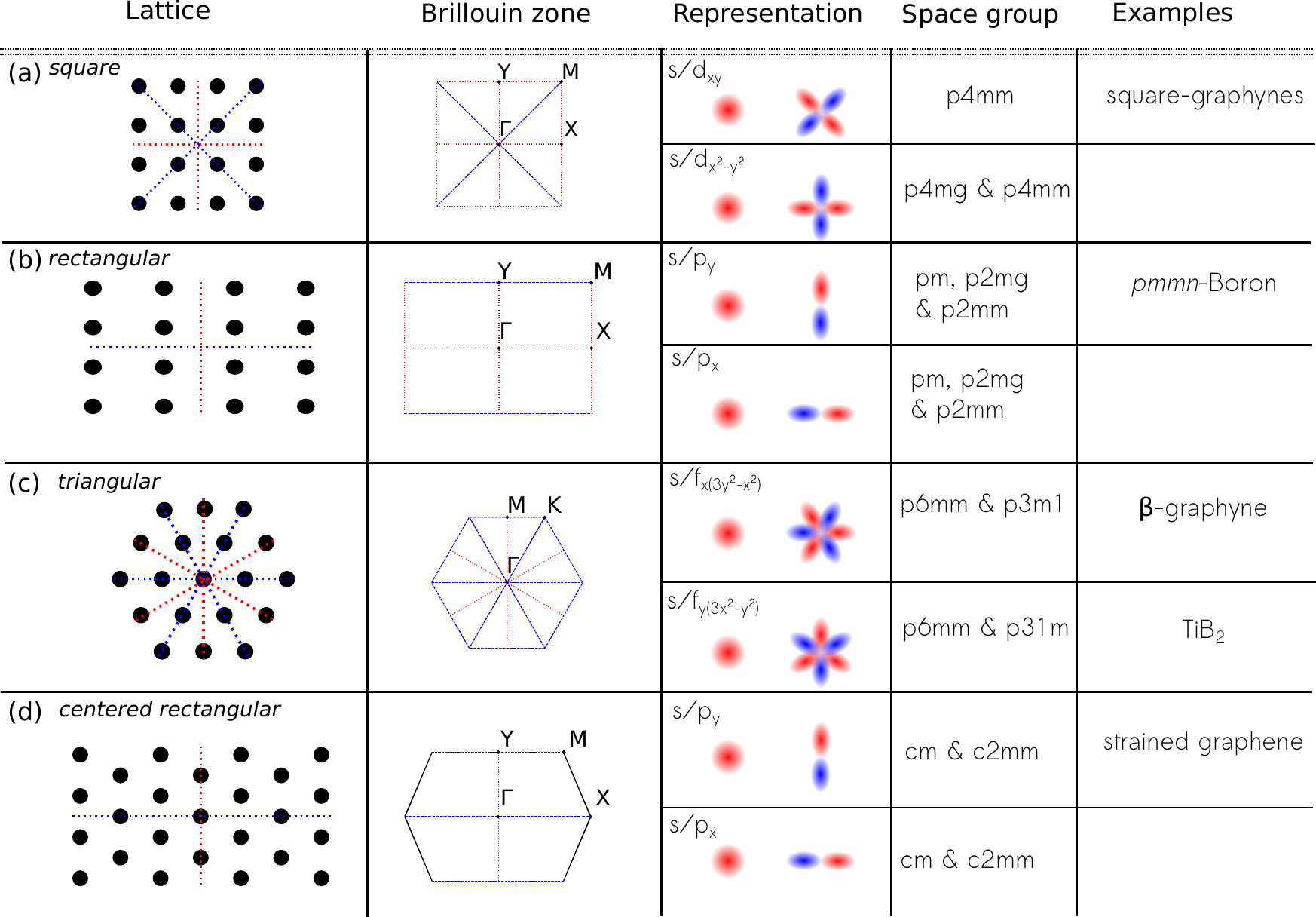}

\caption{\label{fig:lattices} (Color online) Table displaying different realizations of Dirac system with cones along high-symmetry lines. The first column shows four different Bravais lattices. Here, the mirror-symmetry is indicated by the red and blue dashed lines. The corresponding high-symmetry lines are displayed in the Brillouin zone in the second column. For each lattice, we can choose two different representations for the valence and conduction band, which we label by the corresponding atomic orbitals. Furthermore, we denote the different space groups that can host these Dirac cones, and give examples of materials where they occur.}
\label{fig:lattices}
\end{figure*}
\section{Band crossing points}
Dirac systems are characterized by a band-crossing point (BCP), from which the two bands disperse linearly. Such a BCP can be studied using a minimal two-band model, for which the Hamiltonian is given by a Hermitean $2\times 2$ matrix $H_k$, with elements $H_{ij},i,j=1,2$. The presence of a degeneracy means that there exist two eigenvectors $|\Psi\rangle$ and $|\Phi\rangle$ with the same energy $E$. This requires the following relations to hold:
\begin{align}
H_{11}=H_{22},\quad \textrm{Im}[H_{12}]=\textrm{Re}[H_{12}]=0.
\end{align}
Hence, three conditions must be fulfilled for the degeneracy to occur, while there are only two variables, $k_x$ and $k_y$, to tune. This result is known as the Wigner-von Neumann theorem\cite{WvN} (see also Ref.[\onlinecite{gen}] for a generalized Wigner-von Neumann theorem). Therefore, for a BCP an additional symmetry is required, to ensure that at least one of these conditions is automatically satisfied. We will now discuss three different possibilities that may occur. Here and in the remainder of this text, we restrict ourselves to spinless fermions in the presence of time-reversal symmetry (TRS).

(I) In the case of the honeycomb lattice, the BCP occurs at the $K$ and $K'$ points, which are located at the corners of the hexagonal BZ. At these HS points, the symmetry of the lattice ensures that all three conditions are automatically satisfied. This is due to the combination of a threefold rotational symmetry around each lattice point, and the reflection symmetry relating the two sublattices. For this reason, the Dirac cone in graphene is called an essential degeneracy. Other systems with Dirac cones at the $K$ and $K'$ points include the ruby and Kagom\'e lattices.\cite{carmine} Moreover, the Kagom\'e lattice exhibits a quadratic BCP at the $\Gamma$ point.\cite{QBCP}

(II) Let us consider a system that is governed by the Hamiltonian $H$, in the presence of a mirror-symmetry $\hat{R}$. For a momentum $k$ that lies along the HS lines, we have $[\hat{R},H_k]=0$.\footnote{The HS lines corresponding to a mirror-symmetry $\hat{R}$ are composed of momenta for which $\hat{R}k=k$.} Now, by combining this commutation relation with the fact that $\hat{R}^2=I$, where $I$ is the identity, we can block-diagonalize $H_k$
\begin{align}
H_k&=\begin{pmatrix}H_{k,+}&0\\
0&H_{k,-}\end{pmatrix},
\end{align}
where $H_{k,\pm}$ corresponds to the even and odd states, respectively. If we assume that near the Fermi energy one may describe the system  by two bands (even/odd), then the condition for the existence of a Dirac cone along the HS line reduces simply to $H_{k,+}=H_{k,-}$. Hence, the number of conditions matches the number of variables, and there might be Dirac cones in such a system for a finite region in the parameter space of hopping- and on-site energies.

This discussion leads us to the following requirements for constructing a Dirac system with cones along the HS lines: (i) The system should exhibit mirror-symmetry $\hat{R}$; (ii) the two bands closest to the Fermi level should have opposite eigenvalues under the mirror-symmetry $\hat{R}$; (iii) the bands should cross.

Notice that we have not made use of the TRS. However, in the absence of TRS the hopping parameters can pick up a phase, which may break the mirror symmetry.

(III) Finally, we consider a system that exhibits both inversion and TRS. TRS implies that $H_k=H^*_{-k}$, whereas typically the inversion symmetry imposes that $H_k=\sigma_i H_{-k}\sigma_i$, with $i=0,1,$ or $3$, where $\sigma_i$ denote the Pauli matrices and $\sigma_0$ the identity matrix. Hence, if we combine the two relations we obtain $H_k=\sigma_i H^*_k\sigma_i$. In this case, just one of the relations in Eq.~(1) is automatically satisfied, and one might find Dirac cones, because the number of equations matches the number of variables.

Finally, we remark that in the remainder of this paper we limit ourselves to Dirac systems for which the cones are related by symmetry. In particular, this ensures that all Dirac cones lie precisely at the Fermi level. 

\section{2D lattices with cones along high-symmetry lines}
Here, we investigate the various Dirac systems belonging to class II. Given such a system, we may classify it depending on the symmetry of the crystal lattice and the set of HS lines along which the Dirac cones are located. Equivalently, we need to specify the mirror symmetries for which the valence and conduction bands have opposite parity.

The full symmetry of the crystal lattice is given by the space group. A crystal can be seen as a Bravais lattice that encodes the translational symmetry together with a basis that specifies the positions and chemical composition of the atoms in the unit-cell. For example, graphene consists of a two-atom basis of carbon with an underlying triangular Bravais lattice. In general, the crystal corresponds to a decorated Bravais lattice. Hence, the space group that describes the combined symmetry of the Bravais lattice and the basis does not need to coincide with the underlying Bravais lattice. In particular, in two dimensions there are 17 different space groups, see e.g.~Ref.[\onlinecite{dresselhaus}]. Since the class II Dirac systems can only occur for crystals with mirror-symmetry, we will only encounter 9 of them.

Given the space group, we can identify the sets of HS lines and study the corresponding Dirac systems. Moreover, for each realization we construct a minimal Hamiltonian. This Hamiltonian is only valid at low energies around the Dirac cone because away from the Dirac cone other bands may become important. These minimal models can always be chosen to be chiral, meaning that the Hamiltonian is a linear combination of just two Pauli matrices. For each model, we discuss a novel Dirac material as a concrete example. Our analysis focuses on the four primitive Bravais lattices depicted in Fig. 1. Here, we do not consider the oblique lattice, as it does not have any mirror symmetry. Then, for each Bravais lattice we extend the discussion to the space groups that have the same underlying lattice. We discuss the different Bravais lattices in the following order: (i) square, (ii) rectangular, (iii) triangular, and (iv) the centered rectangular. The results of this section are schematically summarized in Fig. 1.

{\bf Square lattice:}\\
A prototypical example of a Dirac system with a square Bravais lattice is square graphyne. Its crystal lattice and band structure are shown, respectively, in Figs.~2(a) and 2(b). Whereas in graphene each unit cell contains two carbon atoms, for square graphyne there are 24 carbon atoms. Its band structure exhibits four highly anisotropic Dirac cones, all located at the boundary of the BZ ($B$ point in Fig.~2(b)). Moreover, along the lines connecting the $\Gamma$ and $M$ points, the band  structure exhibits a small gap. The location of the Dirac cones hints at the fact that the existence of these cones can be understood from the mirror-reflection symmetries of the square lattice.
\begin{figure}[h]
\centering
\includegraphics[width=.45\textwidth]{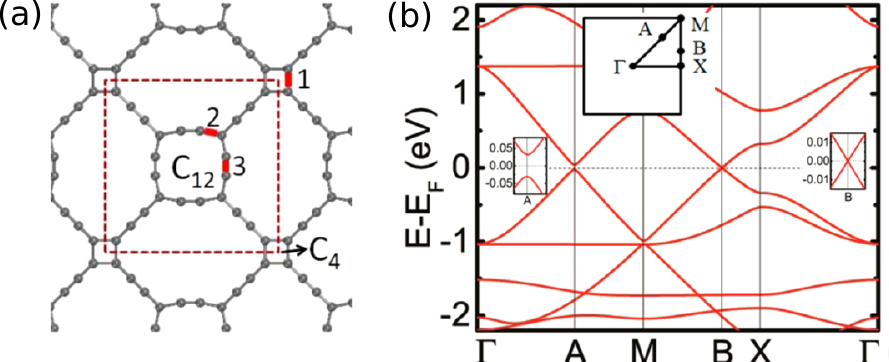}
\caption{ (Color online) (a) Lattice structure of square-graphyne; (b) Band structure along HS lines, obtained from {\it ab initio} calculations. Reprinted (adapted) with permission from Ref.~[\onlinecite{sgraphyne}]. Copyright (2015) American Chemical Society.}
\end{figure}

{\it (i) s and d$_{xy}$}\\
Inspection of Fig.~1(a) shows that the boundary of the BZ together with the lines connecting the $\Gamma-X$ and $\Gamma-Y$ points are associated to the mirror symmetries indicated by the red dashed lines. In particular, one can realize Dirac cones along the red dashed lines by constructing a two-band model composed of two Bloch waves, of which one is even with respect to all mirror-reflection symmetries, and the other is odd with respect to the symmetries indicated by the red dashed lines, but even with respect to the remaining ones. This is achieved, for example, in a two-band model composed of an $s$ and $d_{xy}$ orbital. The minimal Hamiltonian for such a system is given by
\begin{align}
H&=\sum_{k}g(k)(s^\dagger_ks_k-d^\dagger_kd_k)+\sum_{k}h(k)s^\dagger_kd_k+h.c.,
\end{align}
with $g(k_x,k_y)=\epsilon_0+V_{nn}[\cos(k_x)+\cos(k_y)]$ and $h(k_x,k_y)=V_{sd}[\cos(k_x+k_y)-\cos(k_x-k_y)]$. Here, $\epsilon_0$ is an on-site energy, $V_{nn}$ is the nearest-neighbor hopping parameter, and $V_{sd}$ accounts for the hybridization among next-nearest neighbor $s$ and $d$ orbitals. In particular, this model exhibits Dirac cones if $|\epsilon_0|<2|V_{nn}|$.

Note that one does not require specifically atomic $s$ and $d_{xy}$ orbitals; this only illustrates under which representation the bands should transform. For example, in square-graphynes the bands are composed of the atomic $p_z$ orbitals. 

{\it (ii) s and d$_{x^2-y^2}$}\\
To realize Dirac cones along the blue-dashed lines connecting the $\Gamma-M$ points, the Bloch waves should transform under the same representation as the $s$ and $d_{x^2-y^2}$ orbitals, such that they have opposite parity with respect to the symmetries indicated by the blue-dashed lines in Fig. 1. To our knowledge, there are no known materials that realize this model, although it may be promptly engineered using cold atoms. Here, the minimal Hamiltonian with the desired properties is still given by Eq.~(3), but now one has to replace the $d_{xy}$ orbital by a $d_{x^2-y^2}$ orbital. As a result, the hybridization is between nearest-neighbor $s$ and $d$ orbitals and $h(k_x,k_y)=V_{sd}[\cos(k_x)-\cos(k_y)]$. This model also exhibits Dirac cones  for $|\epsilon_0|<2|V_{nn}|$. In addition, the full symmetry of the square Bravais lattice is not required, as this model can also be realized for the space group p4mg.

{\bf Rectangular lattice:}\\
Rectangular crystals with Dirac cones are pervasive; some recent examples that have attracted much attention include multilayer phosphorene, pmmn Boron, and $6,6,12$-graphyne. Whereas pmmn Boron and biased multilayer phosphorene exhibit a single pair of Dirac cones,\cite{zung,phos} the $6,6,12$-graphyne has two pairs. Because these two pairs of Dirac cones are not related by any symmetry, we will not discuss $6,6,12-$graphyne here. In phosphorene, the Dirac cones actually arise due to the presence of glide-reflection symmetry. Therefore we defer its discussion to Sec.~V.

The rectangular lattice exhibits mirror-reflection symmetry in the vertical and horizontal direction, which is represented, respectively, by the red- and blue-dashed lines in Fig.~1(b). For the rectangular Dirac system pmmn Boron (see Figs.~3(a) and 3(b), where the lattice and band structure are depicted, respectively), the Dirac cones are located along the HS line connecting the $\Gamma-X$ points. This indicates that these Dirac cones are a consequence of the mirror-symmetry in the vertical direction.
\begin{figure}[t]
\centering
\includegraphics[width=.45\textwidth]{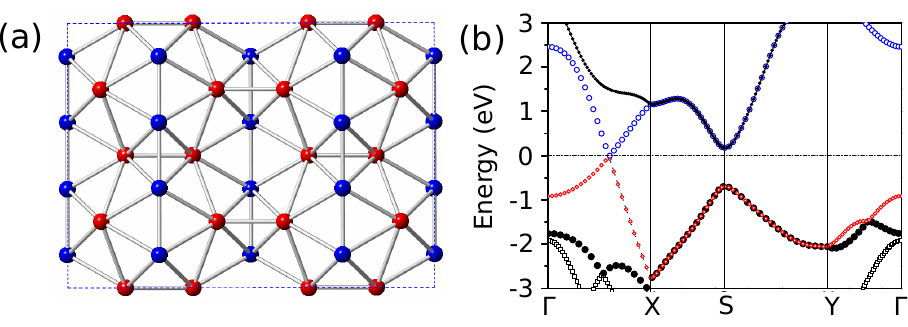}
\caption{ (Color online) (a) Lattice structure of pmmn Boron; (b) Band structure along HS lines, obtained from {\it ab initio} calculations. Figures extracted and edited from Ref.~[\onlinecite{pmmnB}].}
\end{figure}

{\it (i) s and p$_x$ (p$_y$)}\\
The HS lines connecting the $\Gamma-X(Y)$ points and the $Y(X)-M$ points are associated to the mirror-symmetry indicated by the blue (red) dashed lines in Fig.~1(b). A two-band model with Dirac cones along these HS lines should be composed of Bloch waves with opposite parity under reflection in the $y(x)$-direction. In particular, this is realized in a two-band model composed of $s$ and $p_x$($p_y$)-orbitals, for which the minimal Hamiltonian reads
\begin{align}
H&=\sum_{k}g(k)(s^\dagger_ks_k-p^\dagger_kp_k)+\sum_{k}h(k)s^\dagger_kp_k+h.c.,
\end{align}
with $g(k_x,k_y)=\epsilon_0+V_{nn,x}\cos(k_x)+V_{nn,y}\cos(k_y)$, and $h(k_x,k_y)=i V_{sp} \sin(k_x)$ or $h(k_x,k_y)=i V_{sp} \sin(k_y)$. Here, $V_{nn,x(y)}$ denotes the hopping parameter in the $x(y)$ direction, and $V_{sp}$ is the nearest-neighbor hopping parameter that hybridizes the $s$ and $p$ orbitals. This band structure exhibits Dirac cones for $|\epsilon_0+V_{nn,x}|<|V_{nn,y}|$ and $|\epsilon_0-V_{nn,x}|<|V_{nn,y}|$. Note that this Dirac system can be realized in the absence of spatial inversion symmetry, as it can occur for the space groups pm, p2mg, and p2mm. Here, we have assumed that the translation vectors of the rectangular lattice are still given by the unit vectors pointing in the $x$ and $y$ directions.

{\bf Triangular lattice:}\\
Most of the already synthesized two-dimensional materials have a triangular Bravais lattice. In particular, this applies to graphene, silicene, germanene, and stanene. However, for these systems the Dirac cone is an essential degeneracy. A typical example of a material with a triangular lattice and Dirac cones along HS lines is $\beta-$graphyne, see Fig.~4(a). Its unit cell consists of 18 carbon atoms, and has the shape of a hexagon. This carbon allotrope has six inequivalent Dirac cones located along the lines $\Gamma-M$ (see Fig.~4(b)). A different example is provided by TiB$_2$ (see Fig.~5(a)). This material has also six Dirac cones, but they are located along the HS lines $\Gamma-K$ and $\Gamma-K'$ (see Fig.~5(b)). Each of these systems can be seen as a typical example of a triangular Dirac system.
\begin{figure}[t]
\centering
\includegraphics[width=.45\textwidth]{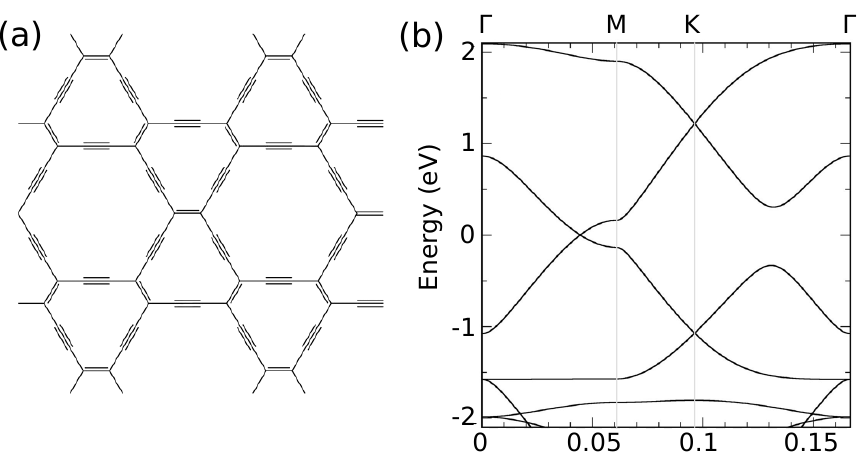}
\caption{ (Color online) (a) Lattice structure of $\beta-$graphyne; (b) Band structure along HS lines, obtained from {\it ab initio} calculations. Figures extracted and edited from Ref.~[\onlinecite{malko2012}].}
\end{figure}

{\it (i) s and f$_{x(3y^2-x^2)}$}\\
From Fig.~1(c), we infer that the HS lines connecting  the $\Gamma-M$ points are associated with the mirror-reflection symmetry indicated by the red-dashed lines. To realize Dirac cones along this set of HS lines, one should construct a two-band model for which the Bloch waves have equal (opposite) parity with respect to the blue (red) dashed lines.  This can be realized in a model composed of $s$ and  $f_{x(3y^2-x^2)}$ orbitals, for which the minimal Hamiltonian reads
\begin{align}
H&=\sum_k g(k)(s^\dagger_ks_k-f^\dagger_kf_k)+\sum_{k}h(k)s^\dagger_kf_k+h.c.,
\end{align}
with $g(k_x,k_y)=\epsilon_0+V_{nn}[\cos(k_x)+\cos(k_x/2+k_y\sqrt{3}/2)+\cos(k_x/2-k_y\sqrt{3}/2)]$ and $h(k_x,k_y)=i V_{sf}[\sin(k_x)+\sin(-k_x/2-k_y\sqrt{3}/2)+\sin(-k_x/2+k_y\sqrt{3}/2)]$. Here, $V_{sf}$ denotes the nearest-neighbor hopping between $s$ and $f$ orbitals. This Hamiltonian exhibits Dirac cones for $|\epsilon_0+V_{nn}|<2|V_{nn}|$.\\

Note that one does not require the full six-fold rotational symmetry of the triangular Bravais lattice because this model can be realized for the space group p3m1.
 
{\it (ii) s and f$_{y(3x^2-y^2)}$}\\
The remaining set of HS lines connects the $\Gamma-K$ and $\Gamma-K'$ points. These HS lines are associated to the mirror-reflection symmetry indicated by the blue-dashed lines in Fig.~1(c). Now, instead, one requires that the two relevant Bloch waves transform as the atomic $s$ and $f_{y(3x^2-y^2)}$ orbitals. In particular, this is realized in TiB$_2$. The minimal Hamiltonian is also given by Eq. (5), but now $h(k_x,k_y)=i V_{sf}[\sin(k_x3/2+k_y \sqrt{3}/2)+\sin(-\sqrt{3} k_y )+\sin(-k_x3/2+k_y \sqrt{3}/2)]$. Due to the symmetry, the hybridization between $s$ and $f$ orbitals here involves next-nearest neighbor hopping parameters. This minimal model exhibits Dirac cones for $|\epsilon_0+1.75 V_{nn}|<2.25|V_{nn}|$.
\begin{figure}[t!]
\centering
\includegraphics[width=.45\textwidth]{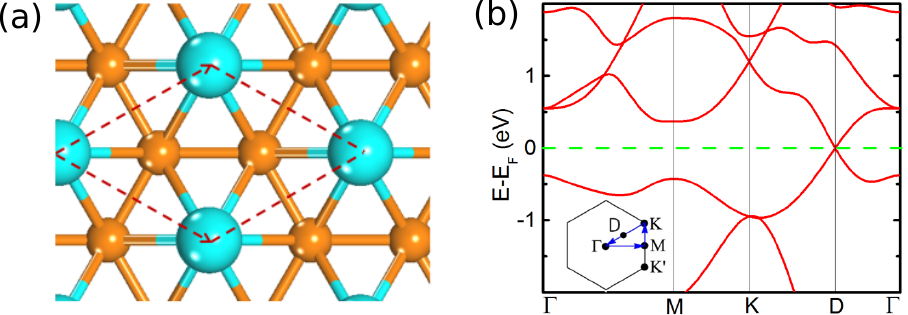}
\caption{ (Color online) (a) Lattice structure of TiB$_2$; (b) Band structure along HS lines, obtained from {\it ab initio} calculations. Figures extracted and edited from Ref.~[\onlinecite{TiB2}].}
\end{figure}

{\bf Centered rectangular}\\
We finalize our analysis with the centered-rectangular lattice. One can obtain this lattice in graphene strained along the armchair direction. The lattice exhibits  mirror-reflection symmetry in the horizontal and in the vertical direction, indicated by the red and blue dashed lines in Fig.~1(d).

{\it (i) s and p$_y$}\\
The HS lines connecting the $\Gamma-X$ and the $Y-M$ points are associated to the vertical mirror reflection symmetry. To construct a model with cones along these HS lines, one should ensure that the Bloch waves have opposite parity with respect to the vertical mirror reflection. This is, for example, realized in a system composed of $s$ and $p_y$ orbitals. In this case, the minimal Hamiltonian is given by
\begin{align}
H&=\sum_k g(k)(s^\dagger_ks_k-p^\dagger_kp_k)+\sum_{k}h(k)s^\dagger_kp_k+h.c.
\end{align}
with $g(k_x,k_y)=\epsilon_0+V_{nn,y}[\cos(k_x/2+k_y\sqrt{3}/2)+\cos(k_x/2-k_y\sqrt{3}/2)]+V_{nn,x}\cos(k_x)$ and $h(k_x,k_y)=i V_{sp}[\sin(k_x/2+k_y\sqrt{3}/2)+\sin(-k_x/2+k_y\sqrt{3}/2)]$. This system may also be realized for the space group cm.
 
{\it (ii) s and $p_x$}\\
The remaining HS line connects the $\Gamma-Y$ points. In this case, the Hamiltonian should be composed of Bloch waves that transform in the same way as the $s$ and $p_x$ orbitals. The corresponding minimal Hamiltonian is still given by Eq.~(6) but now with $h(k_x,k_y)=i V_{sp}\sin(k_x)$. Note that for the centered rectangular lattice we have taken the same lattice vectors as for the triangular lattice.

\section{Away from high-symmetry lines}
The Dirac systems that we have just presented do not form an exhaustive list by any means, as we have limited ourselves to systems with cones along the HS lines. From our analysis in Sec.~II, we know that the systems with Dirac cones away from HS lines must exhibit inversion symmetry. Unless such a system has a single pair of Dirac cones, one needs an additional symmetry that relates the different Dirac cones. In particular, this symmetry can be the rotational symmetry of the lattice, or the combination of rotational and mirror-symmetry. Here, we discuss these cases separately. Moreover, we show how these systems may be seen to descend from class II Dirac systems. 

First, we examine the systems for which the mirror-symmetry is broken, but the rotational symmetry is intact. Specifically, one may obtain such a system by adding a mirror-symmetry breaking term to a class II Dirac Hamiltonian. In real materials, this may be the case due to a substrate. There exist three different space groups describing systems without mirror symmetry but with the inversion symmetry, namely P6, P4, and P2. Note that space group P$N$  has $N-$fold symmetry. For each of these space groups, we present an elementary model. The three different Dirac systems are schematically displayed in Fig. 6.
\begin{figure}[b]
\centering
\includegraphics[width=.45\textwidth]{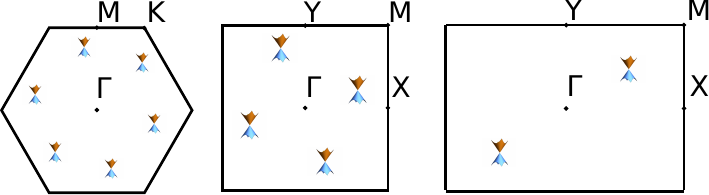}
\caption{ (Color online) The class III Dirac systems, which do no exhibit mirror symmetry.}
\end{figure}

{\bf P6}\\
Starting from the minimal Hamiltonian given in Eq.~(5), which describes either $\beta-$graphyne or TiB$_2$, we may easily break the mirror symmetry, while preserving the six-fold symmetry. In this way, we obtain a Hamiltonian that interpolates between the two systems, which is given by
\begin{align}
H&=\sum_k g(k)(s^\dagger_ks_k-f^\dagger_kf_k)+\sum_{k}h(k)s^\dagger_kf_k+h.c.,
\end{align}
with $g(k_x,k_y)=\epsilon_0+V_{nn}[\cos(k_x)+\cos(k_x/2+k_y\sqrt{3}/2)+\cos(k_x/2-k_y\sqrt{3}/2)]$ and $h(k_x,k_y)=i V_{sf,1}[\sin(k_x)+\sin(-k_x/2-k_y\sqrt{3}/2)+\sin(-k_x/2+k_y\sqrt{3}/2)]+i V_{sf,2}[\sin(k_x3/2+k_y \sqrt{3}/2)+\sin(-\sqrt{3} k_y )+\sin(-k_x3/2+k_y \sqrt{3}/2)]$. For $V_{sf,1}=0$ or $V_{sf,2}=0$, we retrieve the mirror-symmetric models. Depending on the sign of $V_{sf,2}/V_{sf,1}$, the Dirac cones are either shifted to the left or right with respect to the original location.
 
{\bf P4}\\
In a similar fashion, we can smoothly deform the $s/d_{xy}$ and $s/d_{x^2-y^2}$ (see Eq.~(3)) models into each other. Then, the Hamiltonian reads
\begin{align}
H&=\sum_{k}g(k)(s^\dagger_ks_k-d^\dagger_kd_k)+\sum_{k}h(k)s^\dagger_kd_k+h.c.,
\end{align}
with $g(k_x,k_y)=\epsilon_0+V_{nn}[\cos(k_x)+\cos(k_y)]$ and $h(k_x,k_y)=V_{sd,1}[\cos(k_x)-\cos(k_y)]+V_{sd,2}[\cos(k_x+k_y)-\cos(k_x-k_y)]$.

{\bf P2}\\
Finally, we would like to discuss the least symmetric Dirac system, which only has a two-fold symmetry, as encountered in $\alpha$ET$_2$I$_3$. Whereas the two previously discussed systems can only exist on a triangular and square Bravais lattice, respectively, this Dirac system can be hosted on any Bravais lattice. Hence, this is the most robust Dirac system, for it only requires the presence of spatial-inversion symmetry. A minimal model defined on the rectangular Bravais lattice reads
\begin{align}
H&=\sum_{k}g(k)(s^\dagger_ks_k-p^\dagger_kp_k)+\sum_{k}h(k)s^\dagger_kp_k+h.c.,
\end{align}
with $g(k_x,k_y)=\epsilon_0+V_{nn,x}\cos(k_x)+V_{nn,y}\cos(k_y)$ and $h(k_x,k_y)=i V_{sp,1} \sin(k_x)+i V_{sp,2} \sin(k_y)$. By putting $V_{sp,1}=0$ ($V_{sp,2}=0$), we recover the $s/p_x$ ($s/p_y$) models discussed in Sec. III.

These three Dirac systems can be seen to descend from the more symmetric class II systems. Yet, this is not true for all the class III Dirac systems. In particular, $\beta-$graphyne, in the presence of Rashba spin-orbit coupling, exhibits twelve Dirac cones.\cite{PRBR,PRB} Moreover, proposals have been made to realize rectangular optical lattices with four Dirac cones located away from HS lines.\cite{PRA}
\begin{figure}[b!]
\centering
\includegraphics[width=.45\textwidth]{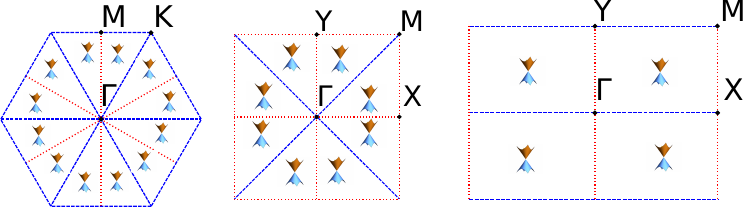}
\caption{ (Color online) The class III Dirac systems, with rotational and mirror symmetry.}
\end{figure}

The conditions to realize these systems are very restrictive, because both inversion and mirror symmetry are required. The inversion symmetry protects the BCP, whereas the mirror symmetry guarantees that all Dirac cones are related. Specifically, for the space groups p6mm, p4mm, and p2mm one may obtain a system with twelve, eight, or four symmetry related Dirac cones, see Fig.~7. Now, we  discuss possible realizations of these systems for each of the space groups.

{\bf P6mm}\\ 
It is possible to obtain a system with twelve symmetry related Dirac cones in several ways. The example of $\beta-$graphyne shows that one may start from six spin-degenerate Dirac cones along HS lines, and then use the Rashba spin-orbit coupling to split these into twelve Dirac cones. However, one does not need to use the spin degree of freedom. In particular, one might realize such a system in twisted bilayers of class II systems. If the two lattices are commensurate, a Moir\'e pattern is formed, see Fig.~8(a).\cite{Moire} To describe such a bilayer, one has to enlarge the unit cell. Correspondingly, the new BZ is reduced in size, see Fig.~8(b). More importantly, the original Dirac cones are not located along HS lines anymore. In particular, this applies to $\beta-$graphyne and TiB$_2$. Furthermore, the inter-layer coupling may shift the position of the Dirac cones. This construction does not depend on any peculiarities of the triangular lattice, therefore, this can easily be extended to the square and rectangular Bravais lattices.
\begin{figure}[t]
\centering
\includegraphics[width=.48\textwidth]{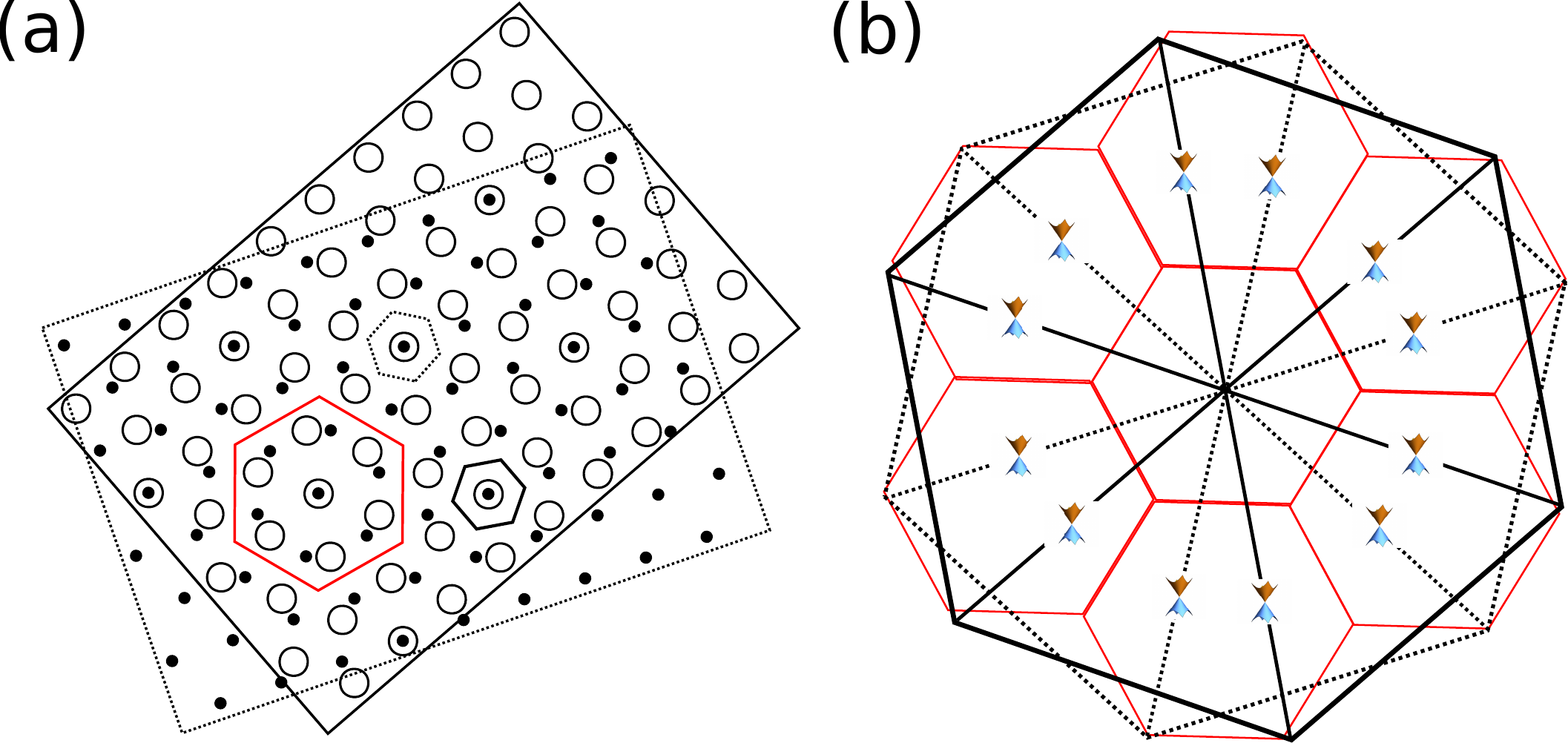}
\caption{ (Color online) (a) A Moir\'e pattern is formed in a twisted bilayer. The original unit cells for the triangular lattices are denoted by the smaller solid and dashed hexagons. The red hexagon denotes the increased unit cell; (b) Here, the dashed (solid) BZ corresponds to the lattice formed by the filled (empty) circles. The reduced BZ is represented by the red central hexagon. We have drawn also the neighbor hexagons to illustrate how these are folded. In this example, we have assumed that the Dirac cones were originally located along the HS lines connecting the $\Gamma-K(K')$ points. }
\end{figure}

{\bf P4mm and P4mg}\\
For the square lattice, it is possible to obtain a system with eight Dirac cones. The previous discussion shows that this may be realized in twisted-bilayer-square graphynes. However, this is not the only way to obtain eight symmetry-related Dirac cones. To see this, we start from a square lattice that is described by the $s/d_{xy}$ model, for which the Dirac cones are located at the boundary of the BZ, as in square-graphynes. Now, we imagine for a moment that we are studying two uncoupled square lattices that are arranged as in Fig.~9(a). Then, if the filled and empty circles are equivalent, the unit cell is decreased by half, and correspondingly the BZ is doubled. The new BZ features eight distinct Dirac cones, located away from HS lines (see Fig.~9(b)). The inclusion of a coupling between the two square lattices moves these cones away from their original position. Precisely this situation can be realized in bilayer-square graphynes, stacked in such a way that the smaller squares (C$_4$ in Fig.~2) are on top of the larger squares (C$_{12}$ in Fig.~2). Note that such a construction does not apply to the triangular lattice, as there one has to triple the BZ, and hence the number of Dirac cones.

{\bf P2mm, P2mg, and C2mm}\\
For the rectangular lattice, one may easily obtain a system with four Dirac cones. For example, one may start from a square lattice with four Dirac cones along the HS lines connecting the $\Gamma-M$ points. Then, if one transforms the square lattice into a rectangular one, while preserving the mirror symmetries of the rectangular lattice in the $x$ and $y$ direction, one keeps the four Dirac cones.
\begin{figure}[t]
\centering
\includegraphics[width=.4\textwidth]{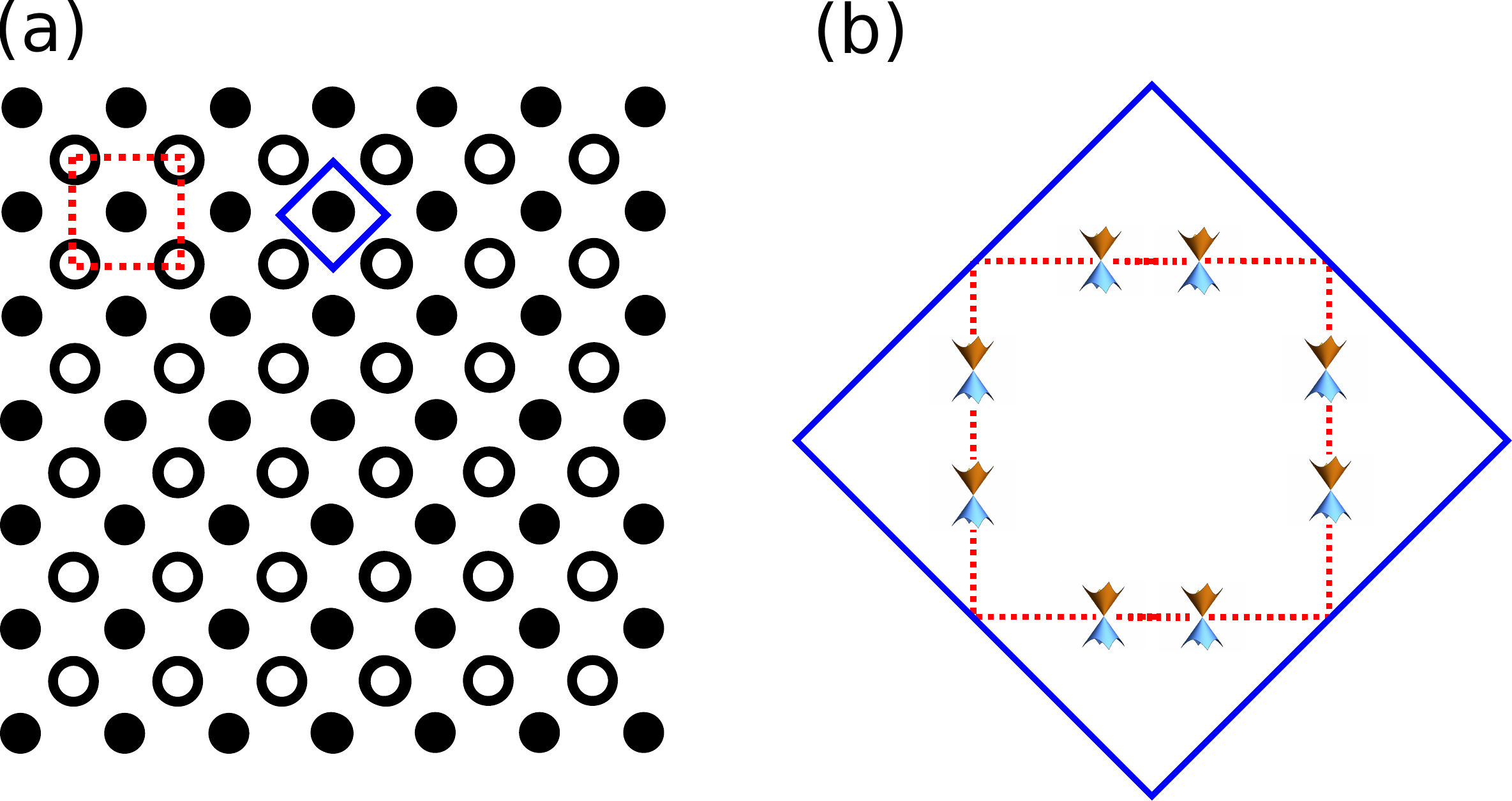}
\caption{ (Color online) Sketch illustrating the doubling of the BZ, and the Dirac cones for the square lattice. (a) The filled and empty dots form a bipartite square lattice. When the filled and empty dots are different (identical), the unit cell is given by the red (blue) square. (b) The red (blue) dashed (solid) square shows the  BZ, corresponding to the larger (smaller) unit cell in (a).}
\end{figure}
\section{Conclusions and discussion}
Dirac systems are very interesting because they are the heart of several unconventional phenomena, like the transparency of graphene or the Klein paradox. They were recently found to occur in materials like graphene, square graphyne or Pmmn boron, which have different lattice structures, ranging from honeycomb to square or triangular, and are composed of different atoms. In this work, we investigate the role of crystal symmetries in Dirac materials. The different systems have been classified within a unified two-band model, which effectively describes the different band structures  that one may obtain for various space groups when one properly accounts for the symmetries of the valence and conduction bands.

We distinguish between three different classes of Dirac systems. Graphene belongs to the first class, and these systems are characterized by a BCP at the HS $K$ and $K'$ points. The class II Dirac systems have their Dirac cones located at HS lines. We have shown that for these systems the Dirac cones occur as a simple consequence of the mirror symmetries of the underlying crystal. This class hosts systems with two, four, and six Dirac cones. The third class contains materials for which the Dirac cones occur at generic points in the BZ. We have shown that these systems can be realized in bilayers composed of class II systems. For example, in bilayer-square graphyne one realizes eight Dirac cones, while this number raises to twelve in twisted-bilayer $\beta-$graphyne.

A careful look at these different classes shows that these systems do not simply differ by the number of Dirac cones and by their position in the BZ, but there are other consequences. For example, in the class I Dirac systems, one can not remove the BCP unless one breaks one of the symmetries. Instead, for the class II and III systems, one may gap out the Dirac cones, whilst preserving all the symmetries. In particular, for class II systems the Dirac cones can annihilate at the time-reversal invariant momenta, whereas it has been shown that for class III systems this can occur at HS lines.\cite{PRBR,PRB}

The classification scheme proposed here relies on certain assumptions. Firstly, we have constrained ourselves to spinless fermions. Hence, we have not touched upon the effects of spin-orbit coupling in these systems. Moreover, we considered time-reversal symmetric systems and described them by generic two-band models. We should stress, however, that this minimal two-band model should be used with caution, especially when higher bands cannot be neglected. Another open question is what happens if we break the TRS. Finally, we did not consider glide-reflection symmetry in our analysis. Nevertheless, our results can be promptly extended to this case. Consider a system described by the Hamiltonian $H$ with glide-reflection symmetry $\hat{G}$. Then, for a momentum $k$ that is invariant under $\hat{G}$ we find $[H_k,\hat{G}]=0$. Since $\hat{G}^2=\exp(i k a)$, with $a$ the lattice constant, the eigenvalues of $\hat{G}$ are not simply $\pm 1$ as for reflection symmetry. Instead, we find the eigenvalues $\pm \exp( ika/2)$, but the bands along the HS lines with opposite eigenvalues still do not repel. The Dirac cones in phosphorene can be understood from such a glide-reflection symmetry.

Our results are relevant to many branches of physics. First of all, one may revert our analysis and use the understanding acquired here to design new Dirac materials. In particular, this approach can be very fruitful in the recently discovered semiconducting nanocrystal superlattices.\cite{Kalesaki} In these systems, the lattice is formed by the nanocrystals, which can be considered as artificial atoms. The advantage then is that one has control over the orbital content and hopping parameters. In this respect, they may serve as the ideal platform to realize the minimal two-band models that we have presented.  An even higher degree of control can be obtained in patterned two-dimensional quantum wells and optical lattices loaded with cold atoms.\cite{pollini} Honeycomb optical lattices have been recently realized,\cite{Sengstock} and experiments impossible in graphene have been viable in these systems. An illustrative example is the merging of Dirac cones upon a deformation of the honeycomb lattice that introduces anisotropies along the different hopping directions.\cite{Goerbig} This anisotropy may be driven by shaking\cite{Koghee} or by superposing different square optical lattices and tuning the relative weight of each of them.\cite{Esslinger}

A very important consequence of the study performed here is that it opens the possibility to realize SU(2N) pseudospin models, which exploit the multiple valley internal degrees of freedom. The possibility to realize optical lattices with high-spin atoms, like Sr or Dy has attracted much attention recently.\cite{highspin} These systems allows one to study SU(2N) magnetism, and were proposed to be a paradigm to engineer synthetic dimensions.\cite{Boada} Here, we show that these concepts can be extended to the valley degree of freedom, with N up to 12.

Finally, we envision that the minimal models can be used in calculations to study the effects of disorder, transport properties, or to study the lattice Green's function beyond the linear dispersion. In addition, these models can be promptly used to study the effects of spin-orbit coupling and to verify whether this leads to topological states with high-spin Chern numbers. Similarly, these models are very suited to study the appearance of edge states in nanoribbons, depending on the termination. Therefore, we hope that our results will provide a motivation to realize these more exotic Dirac systems and trigger further experiments in the field.

\section{Acknowledgments}
We would like to thank Dr. Vladimir Juri\v ci\' c and Dr. Carmine Ortix for fruitful discussions. The authors acknowledge financial support from NWO and the Dutch FOM association with the program "Designing Dirac carriers in semiconductor honeycomb lattices". This work is part of the D-ITP consortium, a program of the Netherlands Organization for Scientific Research (NWO) that is funded by the Dutch Ministry of Education, Culture and Science (OCW).

\end{document}